# Effect of pressure on the superconducting $T_c$ of $MgB_2$

V. G. Tissen*, M. V. Nefedova, N. N. Kolesnikov, M. P. Kulakov

*Institute of Solid State Physics, 142432 Chernogolovka, Moscow Region, Russia*

**Abstract**

Measurements of the superconducting transition temperature $T_c$ have been performed by the inductive method on $MgB_2$ at pressures up to 28 GPa. $T_c$ decreases with applied pressure, exhibiting a cusp at about 9 GPa. We interpret the appearance of this anomaly in the $T_c(P)$ dependence as a result of the pressure-induced electronic transition. Recent band-structure calculations for $MgB_2$ give some support for such an explanation.

*PACS*: 74.62.Fj

*Keywords*: $MgB_2$, superconducting transition temperature, pressure effect.

*Corresponding author. Fax: +7(096)5249701

*E-mail address*: tissen@issp.ac.ru (V. G. Tissen)

## 1. Introduction

The recent discovery of superconductivity in $MgB_2$ with the transition temperature $T_c$ of 39 K [1] has initiated a lot of studies concerning different properties of this compound. Traditionally, the results of the pressure effect on $T_c$ are used to test theoretical models. A linear decrease of $T_c$ with pressure to 1.8, 1.4 and 0.7 GPa at the rates -1.6, -2.0 and -1.11 K/GPa, respectively [2-4], is in a good agreement with the theoretical estimation (-1.4 K/GPa) made within the conventional BCS picture [5].

In the extended pressure range, $T_c$'s of most conventional and high temperature superconductors display a more complicated behavior due to structural or electronic transitions. In the resistivity measurements of $MgB_2$ at pressures up to 25 GPa, two types of the pressure dependences have been revealed: a parabolic one for the samples showing the upturn in the resistance at low temperatures and a linear one for the samples demonstrating the usual metallic behavior [6]. In the latter case, a small deviation from the linearity can be seen at pressures below 10 GPa, although it is almost within the experimental error. Therefore, the authors do not claim any irregularities in the $T_c(P)$ dependences for these samples. On the other hand, the suppression of $T_c$ with the addition of electrons through partial substitution of Al for Mg in $Mg_{1-x}Al_xB_2$ system has been found to be accompanied by a subtle isostructructural transition at $x$=0.1 with a broad two-phase region [7]. Both electron doping and pressure affect $T_c$ in the same direction. Therefore, if the similar structural transition with a step-like decrease in the volume could occur in $MgB_2$ under pressure, this would cause changes in $T_c$, which were more or less spreaded due to pressure inhomogenuity and proximity effects. These circumstances stimulated us to carry out detailed measurements in order to establish if such an anomaly in the $T_c(P)$ dependence of $MgB_2$ does exist.

## 2. Experimental details

Ceramic samples of $MgB_2$ were obtained by direct synthesis from elements. Starting materials were amorphous boron powder and lump pieces of metallic magnesium, both of better than 99.9% purity. These materials, weighed in stoichiometric ratio, were placed, in molybdenum crucible, in medium- pressure furnace with a resistive heater. Under starting Ar-pressure of 12 bar materials were heated up to 1400$^o$C and annealed there for an hour. During the heating, the synthesis of $MgB_2$ is believed to occur at ~900$^o$C. Resulting ceramics was a bronze-color compact dense material with grain size of 6 to 30 microns. Measured density of ceramics was of 2.23 g/cm$^{-3}$. X-ray powder pattern of this ceramics showed the hexagonal $MgB_2$ to be the main constituent ($a_o$=0.3086 nm, $c_o$=0.3520 nm), with admixtures of small quantities of MgO and metallic Mg.

A diamond anvil cell made of nonmagnetic alloy was used to obtain pressures up to 28 GPa. The anvils with 0.5 mm culet diameter are mounted on sapphire backing plates in order to decrease any inductive coupling between the coil system used for ac susceptibility measurements and the metallic surroundings. The sample with dimensions 0.1x0.1x0.03 mm$^3$ and small ruby chips were loaded in the gasket hole filled with methanol-ethanol mixture in the ratio 4:1. The cell was cooled down to 4.2 K in a cryostat. The temperature was measured by a Cu-Fe/Cu thermocouple with an accuracy of about 0.2 K. The pressure was determined at room temperature by ruby fluorescence method. The primary coil yields a magnetic field of 3 Oe at 4.5 kHz. The balanced secondary coils are connected with a lock-in amplifier. The in-phase output signal from the secondary coils was recorded during heating at the rate of about 0.3 K/min. Only the in-phase component has been recorded. The temperature dependent background signal has been appropriately subtracted from the experimental data.

## 3. Results and discussion

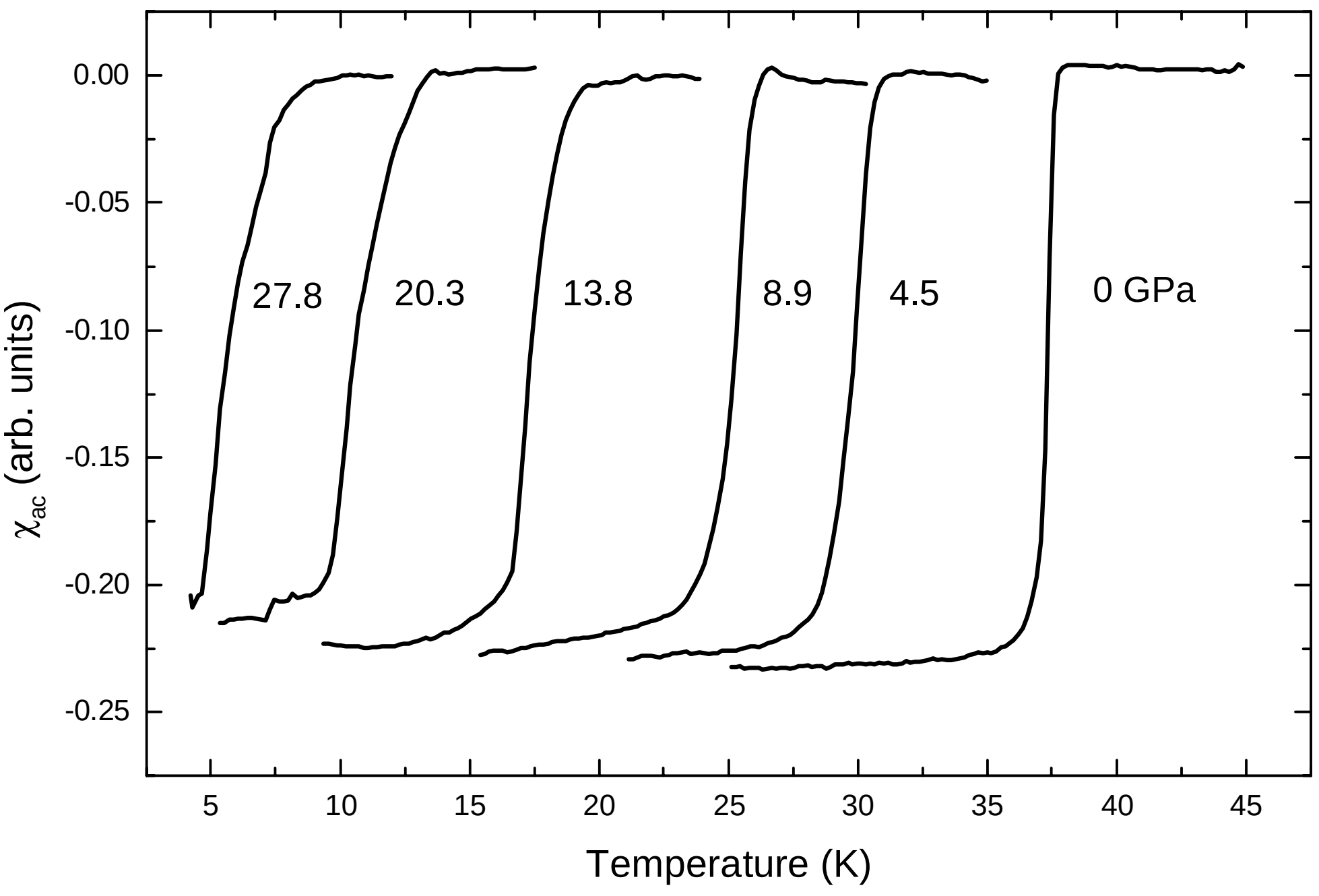

Fig. 1. AC magnetic susceptibility of $MgB_2$ versus temperature at various pressures.

Typical measurements of the ac susceptibility of $MgB_2$ versus temperature at several pressures are displayed in Fig. 1. The superconducting transition is clearly shifted downwards with slight broadening due to increasing nonhydrostaticity with pressure. The pressure dependence of $T_c$, which is defined as the middle of the transition, is shown in Fig. 2. No discontinuity in $T_c$ is observed in the studied pressure range, but a cusp at about 9 GPa can be appreciated. The possible explanation for the appearance of this anomaly will be given below.

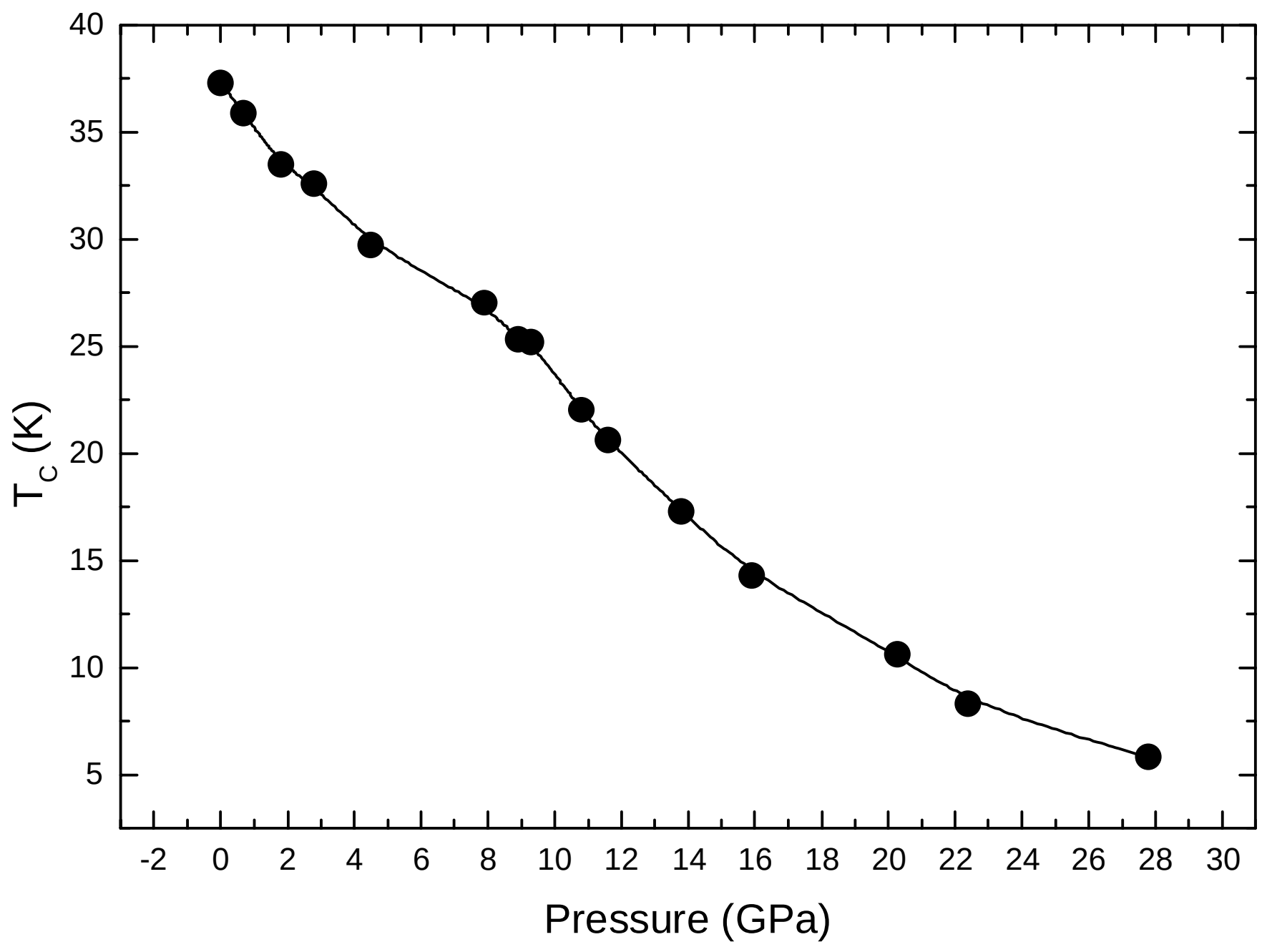

Fig. 2. Superconducting transition temperature (mid-point) as a function of applied pressure.

At ambient pressure $T_c$ was equal to 37.3 K and the initial $\mathrm{d}T_c/\mathrm{d}P$ was estimated to be - 2 K/GPa. At first glance, this pressure coefficient of $T_c$ and those obtained in other experiments (listed in Table 1) are unusually scattered despite the small differences in $T_c$'s. However, as can be seen in Fig. 3, the obvious trend is observed: the higher $T_c$, smaller - $\mathrm{d}T_c/\mathrm{d}P$. Hence, the $T_c$ versus $P$ dependences should exhibit a downward curvature (a curve is located below a tangent), especially for the samples having the highest $T_c$. The samples with the upturn in the resistance behave in such a way in the whole pressure range to 25 GPa [6]. In general, our $T_c(P)$ data (Fig. 2) could be presented as two intersecting curves with the upward curvature. We assume that the downward curvature in $T_c(P)$ could be observable for the present sample in the low-pressure range, that is, there is an inflection point in this range, where are few experimental points. An inspection of the $T_c(P)$ data obtained for the sample with close to ours $T_c$ and $dT_c/dP$ values at pressures to 1.4 GPa [3] allows us to reveal a small but definite deviation from the straight line in accordance with the assumption made above.

Table 1

$T_c$ (the mid-point of the transition) and $dT_c/dP$ values for $MgB_2$.

| $T_c$ (K) | $dT_c/dP$ (K/GPa) | References |
|---|---|---|
| 37.3 | -2 | Present data |
| 37.4 | -1.6 | [2] |
| 37.5 | -1.9 | [3] |
| 38.2 | -1.36 | [8] |
| 39.1 | -1.11 | [4] |
| 39.2 | -1.07 | [9] |

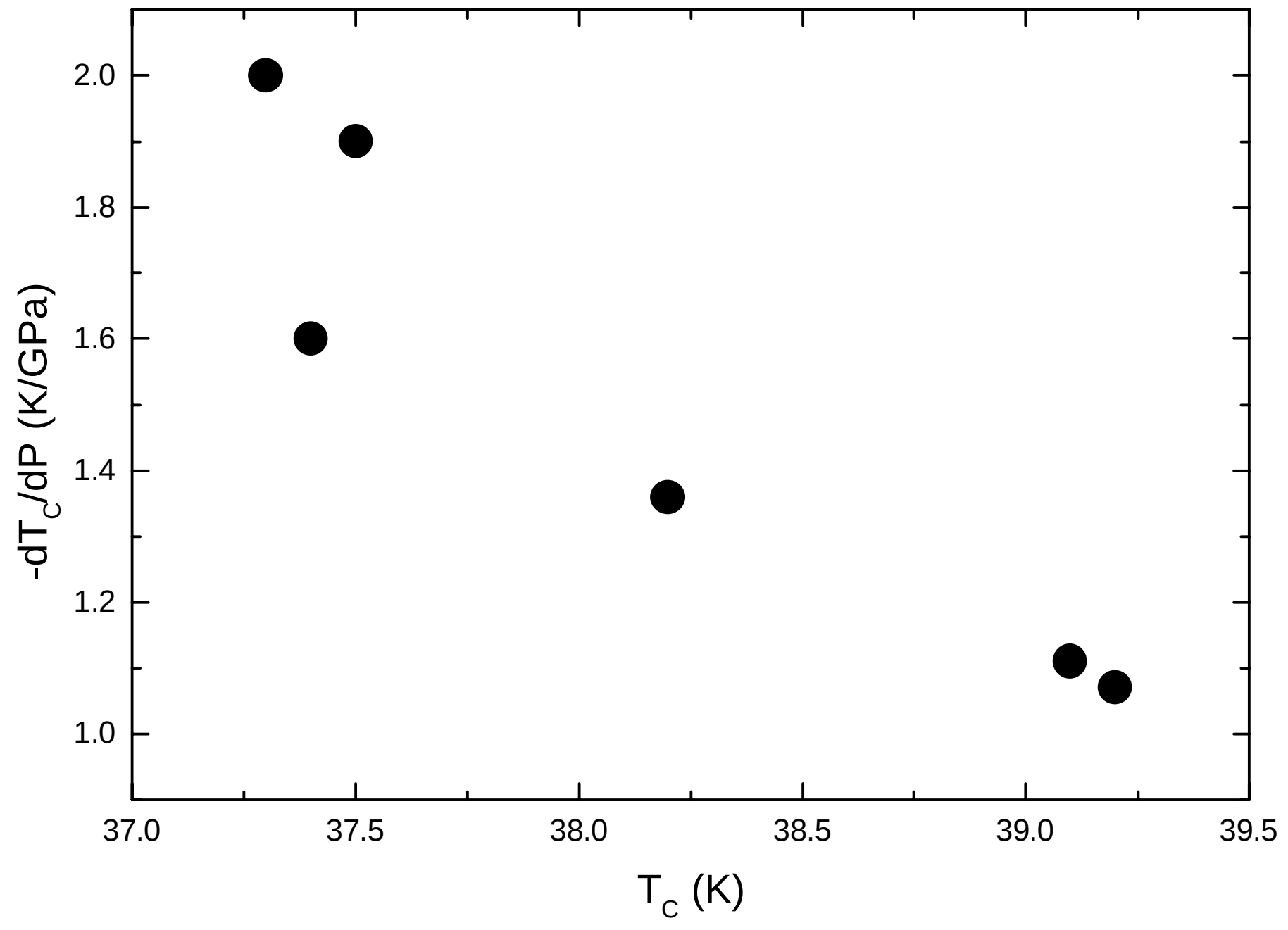

Fig. 3. Pressure derivative of $T_c$ versus $T_c$.

The interesting feature in the $T_c(P)$ dependence is the presence of a cusp, which may originate from the pressure-induced electronic transition. Superconductivity in $MgB_2$ is broadly believed to be connected with the σ bands lying above the Fermi energy $E_F$ along Γ-A line of the Brilloin zone. Band-structure calculations [5, 10, 11] show that, when the parameters of the lattice are reduced, the redistribution of carriers between σ and π bands occurs and the number of holes in the σ band decreases. The shape of the density of states curve above $E_F$ is quite similar to the present $T_c(P)$ dependence, but the value of pressure needed to shift $E_F$ to the van Hove singularity in the density of states is about 97 GPa and, consequently, too high [11]. However, as has been recently shown, the strong electron-phonon coupling in $MgB_2$ results in splitting the σ bands, which were treated earlier as degenerate along Γ-A line of the Brilloin zone, into two subbands [12]. The top of the lower subband is located slightly above $E_F$, therefore, with applying pressure or electron doping, this subband could be shifted below $E_F$, thereby changing the Fermi surface topology and causing the appearance of the van Hove singularity in the density of states. This should result in the electronic Lifshitz transition with particular anomalies in the bulk modulus [13].

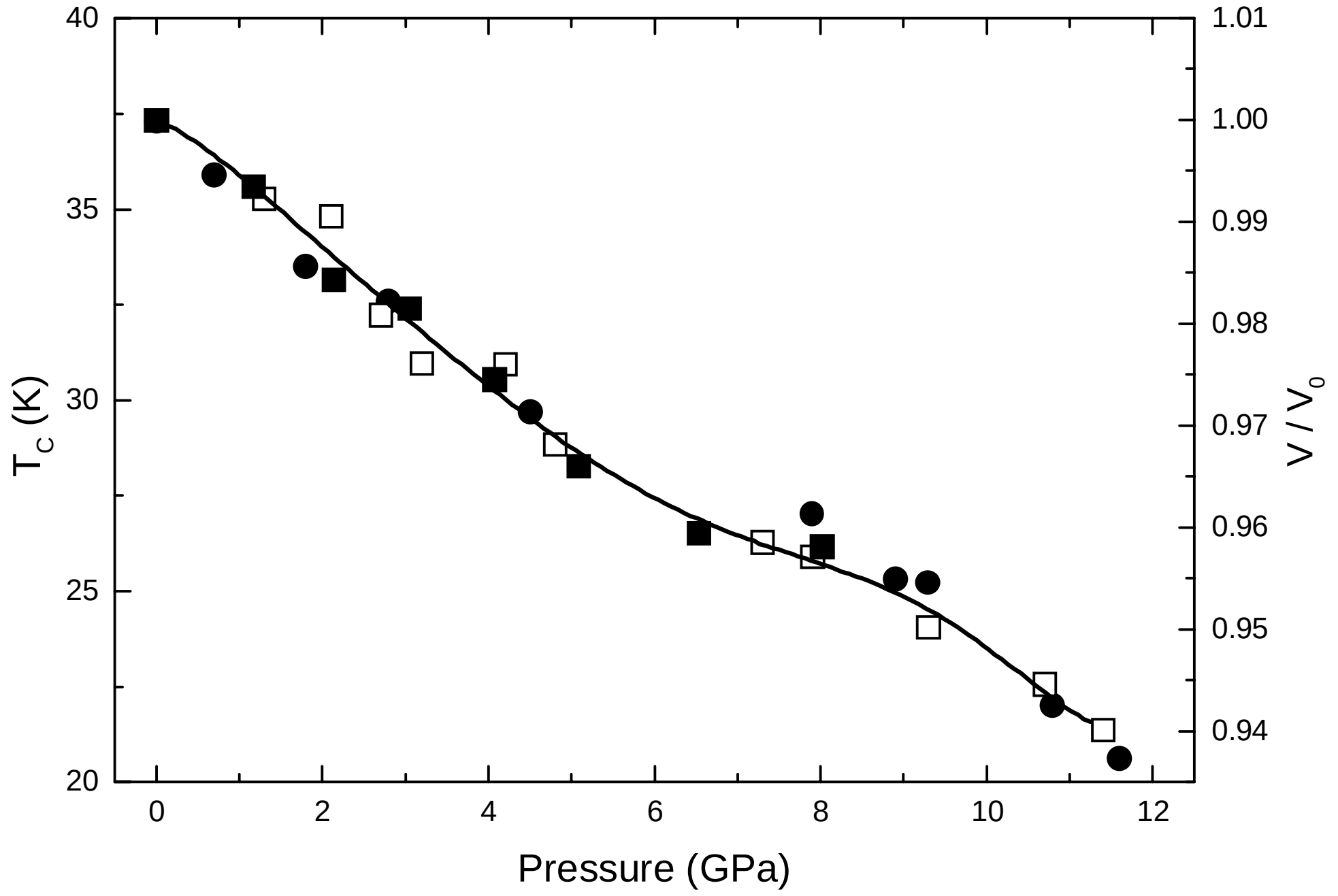

Fig. 4. Pressure dependences of $T_c$ (solid circles) and the relative volume $V/V_0$ (data were taken: solid squares from [14], open squares from [15]). Solid line is a guide to the eye.

Fig. 4 shows the present data for the effect of pressure on $T_c$ of $MgB_2$ (the left axis) together with the results for the relative volume $V/V_0$, which were calculated using the published data for the volume to 8 GPa [14] and for the parameters of the lattice to 12 GPa [15]. As seen in Fig. 4, there is a reasonably good correlation between the anomalies in the two dependences. The bulk modulus softens considerably under pressure above about 9 GPa, but no appreciable drop in the volume is observed. Such a softening may cause an isostructural phase transition if the negative contribution to the bulk modulus is large enough to introduce a van der Waals loop in the pressure-volume curve. This structural transition should occur at higher pressure than needed to shift $E_F$ to the van Hove singularity [13]. One can assume that in $MgB_2$ under pressure, in contrast to the case of the electron doping in $Mg_{1-x}Al_xB_2$ system, the growing positive core contribution to the total bulk modulus, as can be seen in Fig. 4 at pressures below 9 GPa, prevents to develop an isostructural transformation at pressures beyond the electronic transition.

## 4. Conclusions

The pressure dependence of $T_c$ has been determined for $MgB_2$ to 28 GPa. Comparing the present $T_c(P)$ data for $MgB_2$ with the results obtained earlier, we notice that the initial derivative $dT_c/dP$ is strongly dependent on $T_c$ and inversely proportional to it. We assume that the Fermi level crosses the van Hove singularity in the density of states at pressure of about 9 GPa, thereby causing the anomaly in $T_c(P)$. The published results of band-structure calculations and data on $V/V_0(P)$ for $MgB_2$ are in no contradiction with this assumption.

## Acknowledgements

The authors are grateful to E. G. Ponyatovskii for valuable discussions.